\documentclass[aps,pre, reprint, showpacs,superscriptaddress,floatfix,amsmath,amssymb]{revtex4-1}
\usepackage[cm]{fullpage}
\usepackage{float}
\usepackage{graphicx}
\usepackage{bm}
\begin{document}
\newcommand{\sub}[1]{_{\mbox{\scriptsize {#1}}}}


\preprint{AIP/123-QED}

\title[]{Bubble Evolution and Properties in Homogeneous Nucleation Simulations}
\author{Raymond Ang\'elil}
  \affiliation{Institute for Computational Science, University of Zurich, 8057 Zurich, Switzerland}

\author{J\"urg Diemand}
  \affiliation{Institute for Computational Science, University of Zurich, 8057 Zurich, Switzerland}

\author{Kyoko K. Tanaka}
  \affiliation{Institute of Low Temperature Science, Hokkaido University, Sapporo 060-0819, Japan}

\author{Hidekazu Tanaka}
  \affiliation{Institute of Low Temperature Science, Hokkaido University, Sapporo 060-0819, Japan}

\date{\today}
\begin{abstract}
We analyze the properties of naturally formed nano-bubbles in Lennard-Jones molecular dynamics simulations of liquid-to-vapor nucleation in the boiling and the cavitation regimes. The large computational volumes provide a realistic environment at unchanging average temperature and liquid pressure, which allows us to accurately measure properties of bubbles from their inception as stable, critically sized bubbles, to their continued growth into the constant speed regime. Bubble gas densities are up to 50$\%$ lower than the equilibrium vapor densities at the liquid temperature, yet quite close to the gas equilibrium density at the lower gas temperatures measured in the simulations: The latent heat of transformation results in bubble gas temperatures up to 25$\%$ below those of the surrounding bulk liquid. In the case of rapid bubble growth - typical for the cavitation regime - compression of the liquid outside the bubble leads to local temperature increases of up to 5$\%$, likely significant enough to alter the surface tension as well as the local viscosity. The liquid-vapor bubble interface is thinner than expected from planar coexistence simulations by up to $50\%$. Bubbles near the critical size are extremely non-spherical, yet they quickly become spherical as they grow. 
The Rayleigh-Plesset description of bubble-growth gives good agreement in the cavitation regime.  
\end{abstract}


\maketitle

\section{\label{sec:level1}Introduction}
The cavitation and growth of micro-bubbles in superheated or stretched liquids is a common process in nature, and is relevant in many areas of fundamental science, medicine and engineering. Examples include phase changes in the early universe\citep{phase_transitions_early_universe}, the direct detection of dark matter\citep{DMdirect_detection}, targeted drug delivery \citep{drug_delivery}, stimulating accelerated healing of bone fractures\citep{fracture_healing}, cavitation corrosion of water-exposed materials\citep{cavitationDMG,submarine}, volcano fountaining\citep{moar_volcanos,volcanos}, and flavor infusion of plant matter into alcohol in the kitchen\citep{infusion}. However, our understanding of bubble kinetics and dynamics is very limited. This is largely due to the current inability to test various models and frameworks because the properties of the micro-bubbles, such as surface tension\cite{Matsumoto2008}, temperature profiles, density profiles and shapes are not well known\cite{Brennen2013}. This is partially due to the practical limitations of measuring the properties of nano-bubbles under laboratory conditions\cite{Vinogradov2008}. Most molecular dynamics (MD) simulations of bubble nucleation and kinetics that have been done up to now are limited to small computational volumes\cite{Kinjo1998,Wu2003,Novak2007,Sekine2008,Wang2008,Tsuda2008,Abascal2013,Meadley2012}, resulting in sharp pressure increases as bubbles begin
to form\cite{Sekine2008,Tsuda2008}. An option is to barostat the system\cite{Novak2007,Wang2008,Meadley2012}, either by increasing the box size by moving the boundaries, or by rescaling the positions, however, both are unphysical.

\begin{figure*}[]
\includegraphics[scale = 0.65]{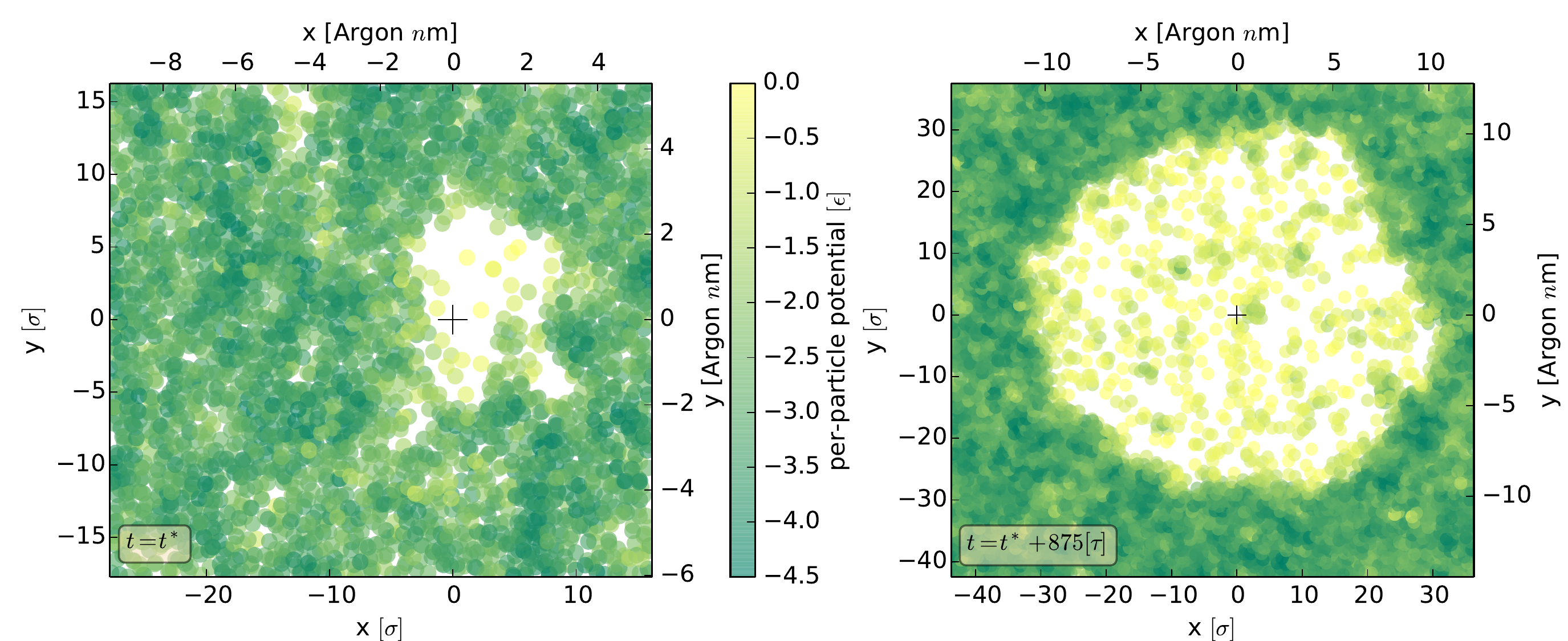}\caption{(Color online) Left: Cross-section through a bubble near the critical size $r_0 = 6.0\,\sigma = 2.0 \,n$m in run T85r3. Particles in a slab of thickness $4\sigma$ in the z-direction are shown and colored according to their potential energy. Right: The same bubble $875\tau$ or $189\,n$s later. The `$+$' markers indicate their center-of-masses. We find incongruous shapes typical for small bubbles.}\label{fig:discs}
\end{figure*}

In this paper we report on the physical properties of bubbles formed from homogeneous nucleation events in Lennard-Jones molecular dynamics simulations in large computational volumes. The large number of particles - half a billion - is necessary to  mitigate finite size effects not found in nature, as well as to ensure realistic bubble growth and high signal-to-noise when measuring bubble nucleation rates and physical properties. To our knowledge, properties of nucleating nano-bubbles have not been described before, and realistic bubble growth curves only once: Just a few sufficiently large MD simulations of bubble nucleation exist\cite{Kuksin2010,Watanabe2010,Watanabe2013,Watanabe2014}, but bubble properties were not reported in these works. Bubble growth curves in the extreme, very fast cavitation regime were analyzed in Kuksin et al. \cite{Kuksin2010}. Our simulations cover the cavitation and the boiling regimes, they initially contain a metastable fluid - a superheated or sub-pressurized Lennard-Jones liquid at chosen temperature and pressure, which is allowed to naturally nucleate bubbles which continue to grow. The simulations we examine here are part of a larger set of runs; whose results regarding nucleation rates, critical sizes, size distributions, and accompanying comparisons to nucleation models and experiments are reported in Diemand et al.\cite{bubble_paper_1}. This paper reports on the physical properties of the bubbles from their spontaneous nucleation near the critical size, to their continued growth by factors of up to 8000 in volume. We measure their temperatures, constituent radial velocities, shapes, densities, and growth rates. We use this information to test the Rayleigh-Plesset bubble growth model paradigm.

\section{\label{sec:numerical_simulations}Numerical Simulations}
The Large-scale Atomic/Molecular Massively Parallel Simulator (or LAMMPS) computer program\citep{lammps} was used to perform the simulations. The pure Lennard-Jones (LJ) potential reads 
\begin{equation}
\frac{u_{\textrm{LJ}} \left(r\right)}{4\epsilon} = \left(\frac{\sigma}{r} \right)^{12} - \left(\frac{\sigma}{r} \right)^6,
\end{equation}
however, we truncate and force-shift it (TSF) so that both the potential and its first derivative vanish at the cut-off radius $r_c$. This is done by setting the interaction potential to
\begin{equation}
u_{\textrm{TSF}}\left(r\right) = u_{\textrm{LJ}}\left(r\right) - u_{\textrm{LJ}}\left(r_c\right) - \left(r-r_c\right)\frac{du_{\textrm{LJ}}\left(r_c\right)}{dr}.
\end{equation}
The properties of the TSF-LJ fluid depends on the cut-off scale. We set ours at $r_c =2.5\sigma,$ as done in recent nucleation simulation studies\citep{wang,escobedo}. In the Argon system, the units are $\sigma = 3.405 \,\text{\AA}$, $\epsilon/k = 161.3 \,\textrm{K}$ and $\tau = 2.16\,\textrm{ps}$. The integration routine is the Verlet integrator, with a time-step either of $\Delta t = 0.0025\tau$ or $\Delta t = 0.004\tau,$ (for the lower temperature runs). The simulations are NVE - constant particle number, volume, and energy. All simulations contain $N=536'870'912$ particles, and cover four temperatures $kT = 0.6, 0.7, 0.8, 0.855 \, \epsilon/k$, at pressures which extend from the cavitation to the boiling regime. In many cases it is useful to compare measured thermodynamic bubble quantities to the gas-liquid coexistence value at the same temperature of the bubble run. To this end, we have performed supplementary slab coexistence simulations. More details on these simulations, including convergence tests, may be found in Diemand et al.\cite{bubble_paper_1}.

At each temperature a number of simulations are performed at different pressures, however we do not analyze the complete set in this paper. In this paper there are two types of simulation-outputted data which we analyze:
\begin{enumerate}
\item \textbf{Bubble size and location data} By overlaying a $\left(3\sigma \right)^3$ cell-length grid on the simulation box, at regular intervals, the positions of all cells with a number density less than $0.2\sigma^{-3}$ are outputted. These may then be linked together by iteratively checking for nearby low density cells. This procedure is computationally cheap and so is done regularly - usually every 1'000 integration time-steps. Under the assumption of sphericity, these voids may be converted into a bubble radius $r$. We used this information to measure nucleation rates and bubble size distributions in \cite{bubble_paper_1}.In this paper we use this information to analyze the bubble growth rates, see section \ref{sec:bubble_growth}.
\item \textbf{Particle data of bubbles and their surroundings} To analyze the physical properties of the bubbles, more information is necessary. For this, we output the positions, velocities, coordination numbers, and potential energies of all the particles within a cubic volume centered around the bubble. This outputted region typically extends deep into the liquid (usually at least a few times larger than the bubble itself), because, as we shall report, only far from the bubble do the liquid properties settle to the bulk expectations. 
\end{enumerate}

Full particle data of bubbles and their surroundings from four simulations are analyzed in this paper, three in the cavitation regime, and one in the boiling:
\begin{itemize}
\item \textbf{T6r2} Our most-negative pressure run. Bubble growth is extremely rapid here, although the nucleation rate is low. We measure growth velocities up to $0.55 \, \sigma/\tau$ ($92\, ms^{-1}$). At the end of the simulation, we analyze the three (and only) bubbles which form. By the end of the run, the liquid pressure has increased by $6\%$.
\item \textbf{T7r2} This negative liquid pressure run saw the nucleation of only a few stable bubbles. We examine its largest bubble at four different times. $100, 200, 300, 400\tau$ after its formation time $t^*$. Up until $300\tau$ after the formation time, the pressure has risen by only $4\%$. By $400$ however, it has halved.   
\item \textbf{T8r2} This run is also in the cavitation regime. We analyze its four largest bubbles at the end of the simulation. The liquid pressure rises by $1\%$ over the entire run. 
\item \textbf{T85r3} In this positive-pressure run we analyze seven bubbles. Three of these at two different times, and the last only at the later time. Over this time the liquid pressure rises by $10\%$. The smallest bubble we analyze in this run has volume approximately equal to the critical bubble volume. This bubble is pictured at two different times in figure \ref{fig:discs}.
\end{itemize}

Further details on the simulations and the relevant thermodynamic variables are given in table \ref{tab:details}.

\begin{table*}\caption{Average simulation temperature $T$, and average liquid pressure $P_l$ over the run, up to and including the pressure increases in later stages. Details on the measurements of the critical bubble sizes $r^*$ and nucleation rates $J$ may be found in Diemand et al. \cite{bubble_paper_1}. $t_{end}$ is the duration of the simulation run. $P_e$ is the equilibrium pressure at the run temperature, $n_v$ the equilibrium vapor number density at the run temperature, $n_l$ the equilibrium liquid number density at the run temperature, and $\gamma$ the planar surface tension at the run temperature. We have calculated the thermodynamic parameters from equilibrium slab simulations.}\label{tab:details}
\begin{tabular}{ c | c c c c c | c  c  c  c }
\hline \hline
Run ID &  $T$           &$P_l$                &  $r^*$       &  $J$                     &   $t_{end}$  &  $P_{e}$               &  $n_v$      & $n_l$      & $\gamma$ 
\\
       &  $[\epsilon/k]$&$[\epsilon/\sigma^3]$&  $[\sigma]$  &  $[\sigma^{-3}\tau^{-1}]$&   $[\tau]$   &  $[\epsilon/\sigma^3]$ &  $[1/\sigma^3]$& $[1/\sigma^3]$& $[\epsilon/\sigma^2]$ 
\\ \hline
\textbf{T6r2}&  $ 0.6   $& $-0.336$            &  $2.9\pm 0.4$&$2.5\pm 1.2\times10^{-11}$&   $328$      &  $ 0.0034$             &  $0.00606$     & $ 0.792$      & $ 0.511$ 
\\ \hline
\textbf{T7r2}&  $ 0.703 $& $-0.162$            &  $3.0\pm 1.0$&$2.0\pm 1.1\times10^{-11}$&   $1300$     &  $ 0.0118$             &  $0.0198$      & $ 0.729$      & $ 0.329$
\\ \hline
\textbf{T8r2}&  $ 0.800 $& $-0.371$            &  $5.2\pm 0.3$&$4.2\pm 2.8\times10^{-11}$   &$437.5$    &  $ 0.0303$              &  $0.0505$      & $ 0.652$      & $ 0.168$
\\ \hline
\textbf{T85r3}&  $ 0.855$& $0.0200$            &  $6.9\pm 0.2$&$2.9\pm 0.6\times10^{-12}$   &$2885$     &  $ 0.0461$              &  $0.0833$      & $ 0.595$      & $ 0.0900$
\\ \hline \hline
\end{tabular}
\end{table*}

Because the theoretical models to which we shall compare our numerical results assume bubbles to be spherical, it is convenient to radially bin measured quantities, even if the bubble is not completely spherical. To do this, we must first choose a center for the spherical coordinate system, which should correspond to the bubble center. The bubble center is chosen to be the average of all particles with a coordination number (for a chosen search distance of $r_s = 1.6\,[\sigma]$) less than or equal to 2. This effectively results in the estimated center positions being most heavily weighted by particles within the inner part of the transition region, as members of the gas typically have zero neighbors, and those of the liquid typically more than 6. Unless otherwise indicated, the radial bins are of size $0.7\sigma.$

Section \ref{sec:interfaces} analyses the density profiles of the bubbles. In section \ref{sec:temperatures} we measure the bubble temperature profiles and bulk motions. Section \ref{sec:shapes} uses principal component analysis to determine the bubble shapes. Finally section \ref{sec:bubble_growth} measured growth velocities and compares them to expectations from hydrodynamics.

\section{\label{sec:interfaces}Density profiles}
\begin{figure}
\includegraphics[scale = 0.5]{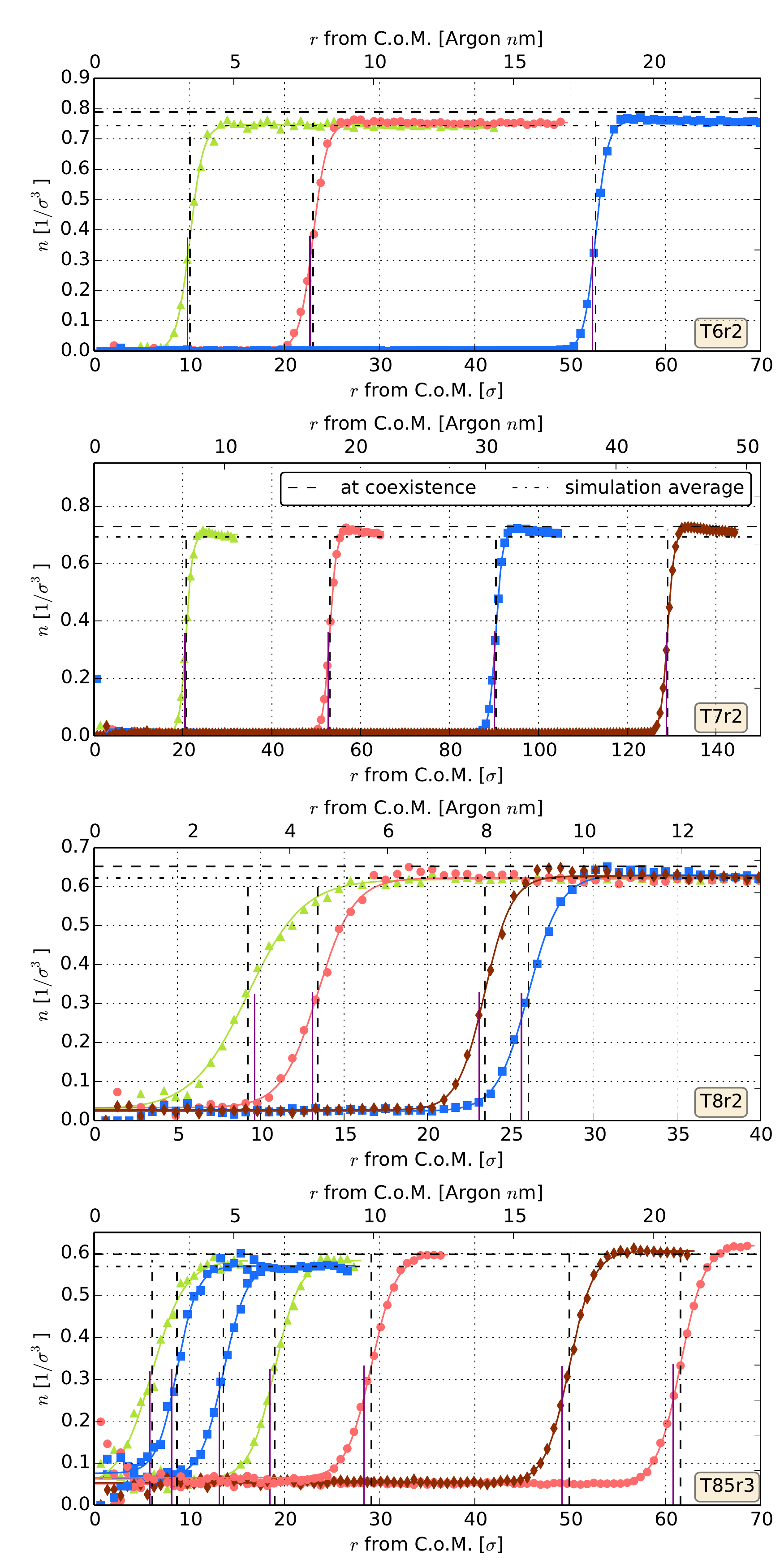}
\caption{(Color online) Bubble density profiles for 3 runs. The dotted vertical lines indicate the center of the transition region according to the fit to equation \eqref{eq:number_density_fit}. The solid purple vertical lines mark the equimolar radii. For the $kT=0.6, 0.7\epsilon$ runs, we note that the densities directly outside the bubble are noticeably higher than the simulation averages, yet they are observed to decrease yet further into the liquid. This is because the rapid bubble growth at low pressures creates a shock-wave around the bubble (see discussion in section \ref{sec:bubble_growth}). The smallest bubbles in T85r3 and T8r2 show significantly larger transition regions than the larger bubbles. This is likely an artificial contribution from their asphericity (see figure \ref{fig:PCA} for principle component shape analysis). Figure \ref{fig:gas_and_d_vs_r0} shows the resulting fit values and associated errors for these bubbles.}\label{fig:number_density_profile}
\end{figure}

For each bubble, we calculate its center and radially bin the particle number density. These are plotted in figure \ref{fig:number_density_profile} for a few bubble examples. For gas-liquid interfaces it is often convenient to fit a fitting function of the form
\begin{equation}\label{eq:number_density_fit}
n\left(r\right) = \frac{1}{2}\left[n_g + n_l -(n_g - n_l)\cdot\left(\tanh \frac{2\left(r-r_0 \right)}{d} \right) \right],
\end{equation}
where $n_g$ and $n_l$ are the inner and outer densities, $r_0$ is the midpoint of the transition region, and $d$ is half the transition region width. Square gradient theory\citep{Wilhelsmen} qualitatively predicts this density profile behavior over the gas-liquid interface.

\begin{figure}
\includegraphics[scale = 0.55]{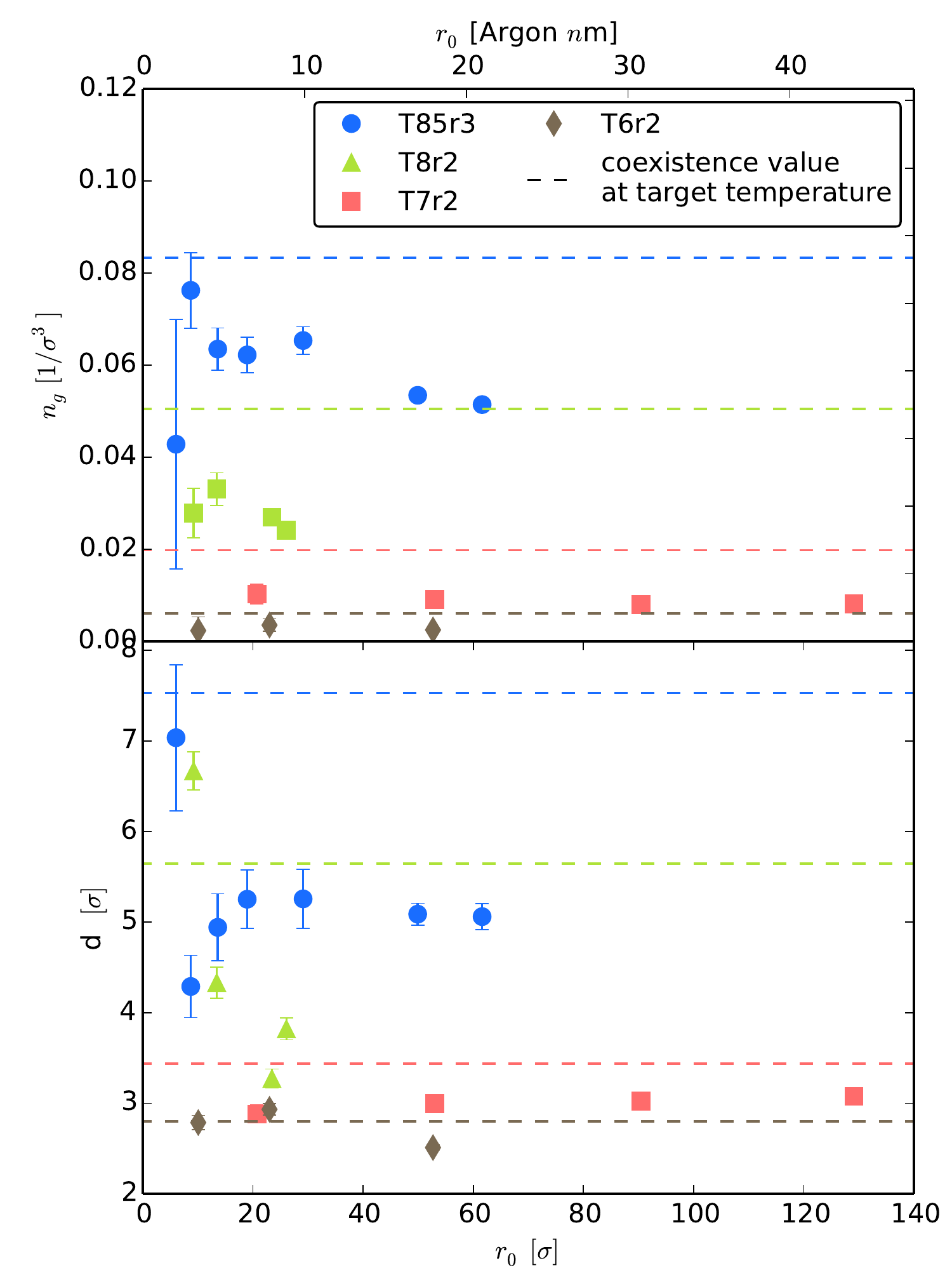}
\caption{(Color online) Gas density $n_g$ (upper panel) and interface transition width $d$ (lower panel) from fitting equation \eqref{eq:number_density_fit} to the bubble profiles in figure \ref{fig:number_density_profile}.
The horizontal axis is the fitted transition region midpoint $r_0$. Over all runs the gas density is between 50\% to 80\% of the coexistence value at the run average temperature. This is likely due to the lower temperatures measured in the bubble gas. A similar discrepancy is evident for the bubble interface widths. The non-isothermality over the transition regions may be responsible for this discrepancy.}\label{fig:gas_and_d_vs_r0}
\end{figure}

Figure \ref{fig:number_density_profile} also shows examples of the fits of equation \eqref{eq:number_density_fit} to the measured binned number densities. These fits provide a useful bubble size definition $r_0$, which we will use in coming sections. The bubble profile fits also allow us to calculate the equimolar radius $R$ of each bubble. Nucleation and bubble kinetic theory are formulated in the sharp-bubble-interface paradigm. When comparisons are made to these theories we therefore use the equimolar radius size $R$. 
Fit results for the gas and liquid  bubble densities and transition widths along with comparisons to bulk values are plotted in figure \ref{fig:gas_and_d_vs_r0}. 

We find that:
\begin{itemize}
\item For bubbles in the cavitation regime, the liquid density directly outside the bubble has a higher density (5\% at $kT = 0.7\epsilon$) than the bulk liquid. This over-density decreases gradually further away from the bubble, indicating the presence of a shock-wave due to the rapid bubble expansion, and the comparatively low sound speed. 
\item The gas density within the bubble is significantly lower than that expected from coexistence simulations at the run temperature. At $kT = 0.7\epsilon,$ the gas density for all bubbles is $50\%$ of the run temperature coexistence value, at $kT = 0.8\epsilon$ between $50-70\%$, and at $kT = 0.855\epsilon,$ between $60-90\%.$ However, we note that the bubble gas densities are close to the equilibrium value at the lower temperature measured for the gas. (See section \ref{sec:temperatures}.)
\item Bubble transition regions are broader in the higher temperature runs than in the lower, however, in all runs, they are lower than the target temperature coexistence values, by typically between $50-80\%$. As above for the gas densities, we again note that the widths are close to the equilibrium value at the temperature of the gas or the interface. (See section \ref{sec:temperatures}.) That higher temperatures result in broader transition regions is expected from Density Functional Theory applied to bubbles[Shen and Debenedetti, 2001]\citep{shenDebenedetti}. 
\item For bubbles sizes $\lessapprox 20\sigma$, we do not have strong enough statistics to make conclusions as to how the radius of curvature affects the transition region width. However, for bubbles $\gtrapprox 20\sigma$, the transition widths have converged to within $5\%$ of their final values. 
\end{itemize}

The non-isothermality of the gas plays a major role in offsetting the measured values from their bulk average run temperature values, which we discuss in the next section. Small bubble curvature is expected to play a more minor role. The density profile, together with the pressure tensor profile provides the surface tension. We have attempted to directly calculate the pressure tensor bubble profiles using the Kirkwood-Irving method\citep{thompson}, however the signal-to-noise is too low to make useful conclusions. 

\section{\label{sec:temperatures}Temperatures and bulk motions}

Bubble nucleation and evolution models typically assume that the bubble formation and growth process is isothermal\citep{blander_katz}, and takes place at the temperature of the liquid. In this section we test this assumption by considering the kinetic energies of the particles.

\begin{figure}
\includegraphics[scale = 0.6]{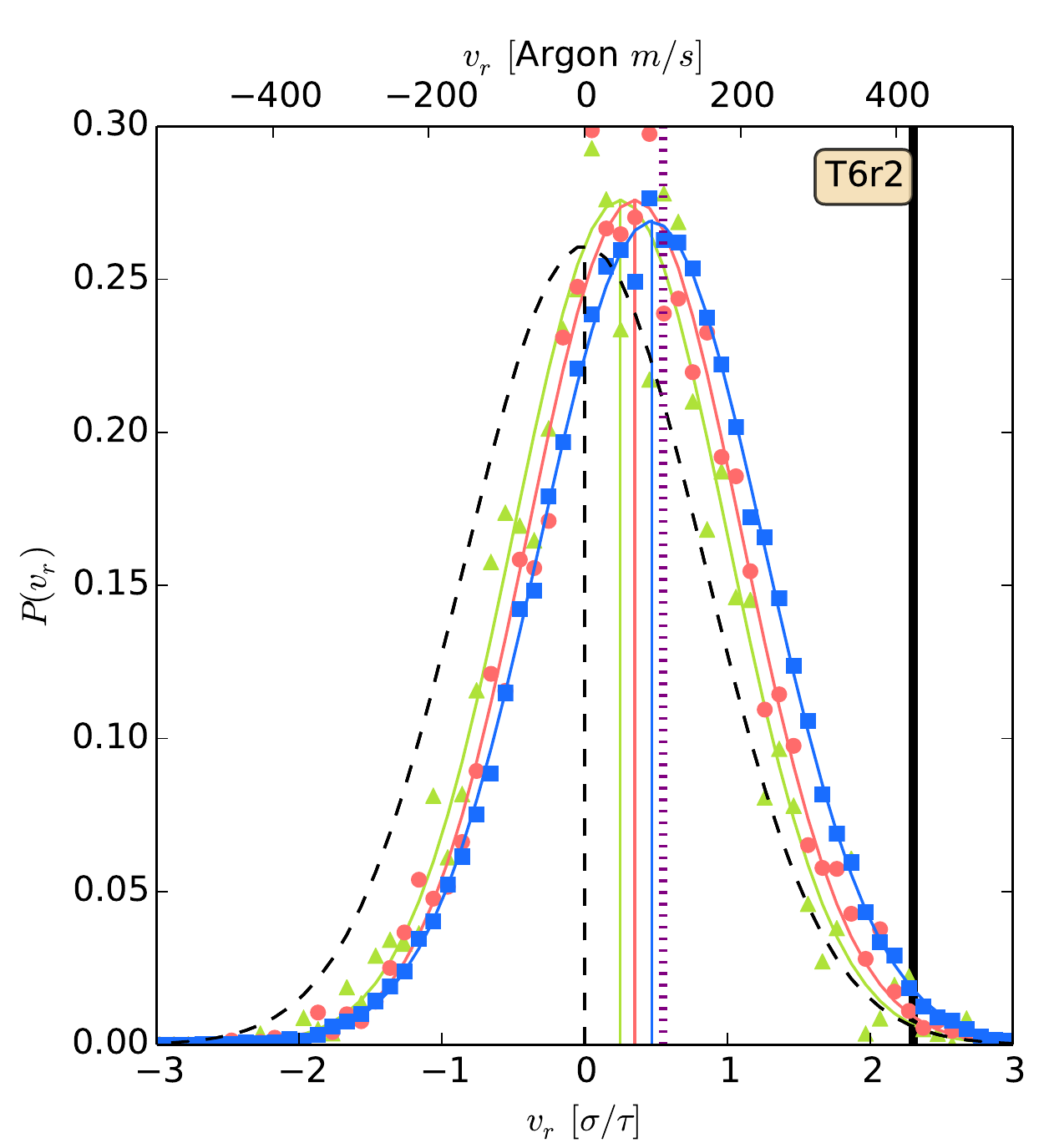}\caption{(Color online) Transition region radial velocity probability distributions for the three bubbles of run T6r2. Only particles within the radial bin covering the transition region are used. The black dotted curve is a reference Gaussian for the average run temperature. Due to the rapid bubble growth in the cavitation regime, the entire distribution is shifted to the right. The thin solid vertical lines indicate the midpoints of the fitted Gaussians, while the dotted purple line indicates the expected bubble linear regime terminal velocity $v_{max} = 0.55\sigma/\tau$ for this run. The thicker vertical black line indicates the sound speed in the liquid.}
\label{fig:interface_rad_vel_t6}
\end{figure}

\begin{figure}
\includegraphics[scale = 0.6]{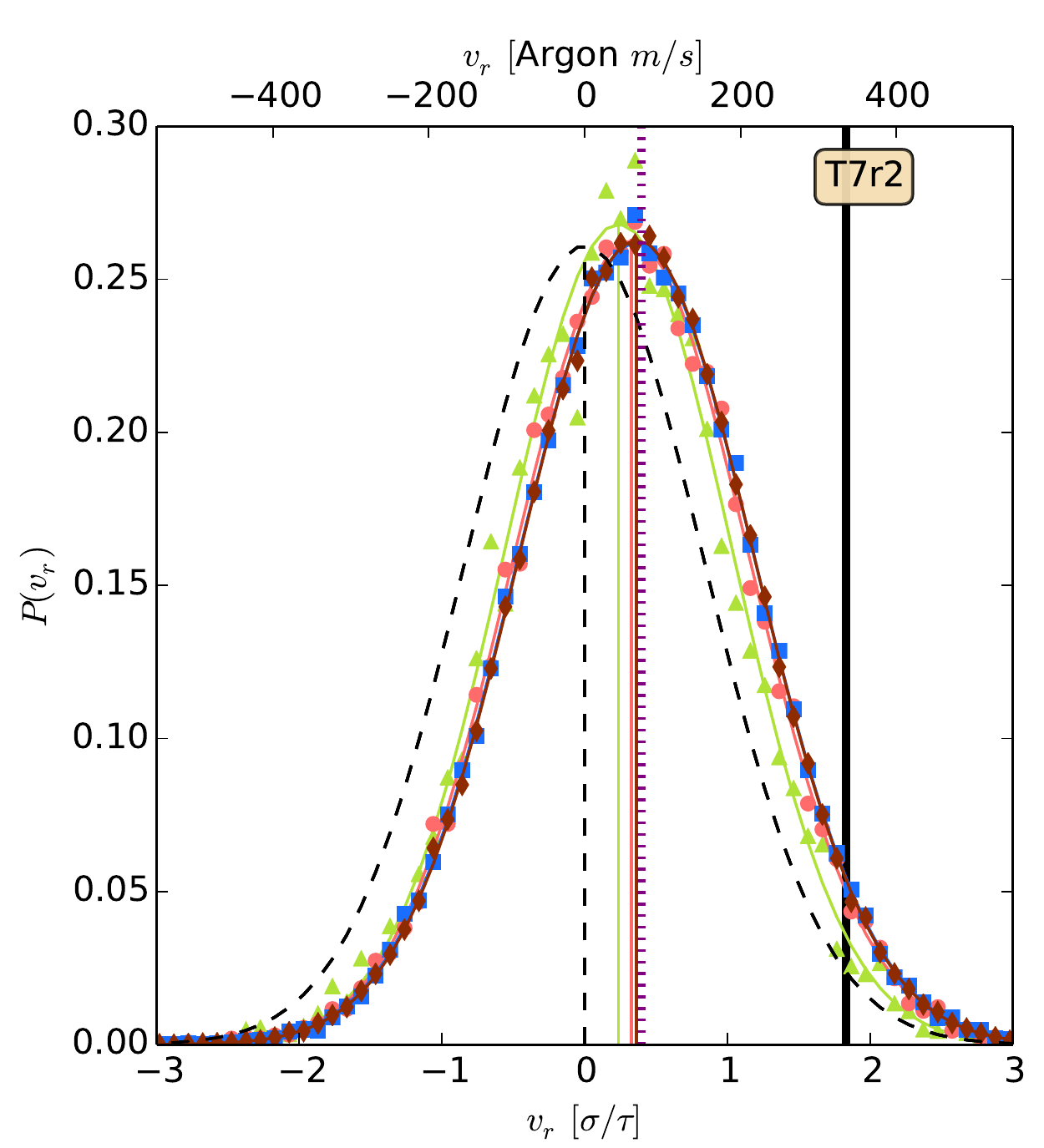}\caption{(Color online) Transition region radial velocity probability distributions for the bubbles in run T7r2. The thick vertical black line indicates the sound speed in the liquid.}\label{fig:interface_rad_vel_t7}
\end{figure}

\begin{figure}
\includegraphics[scale = 0.6]{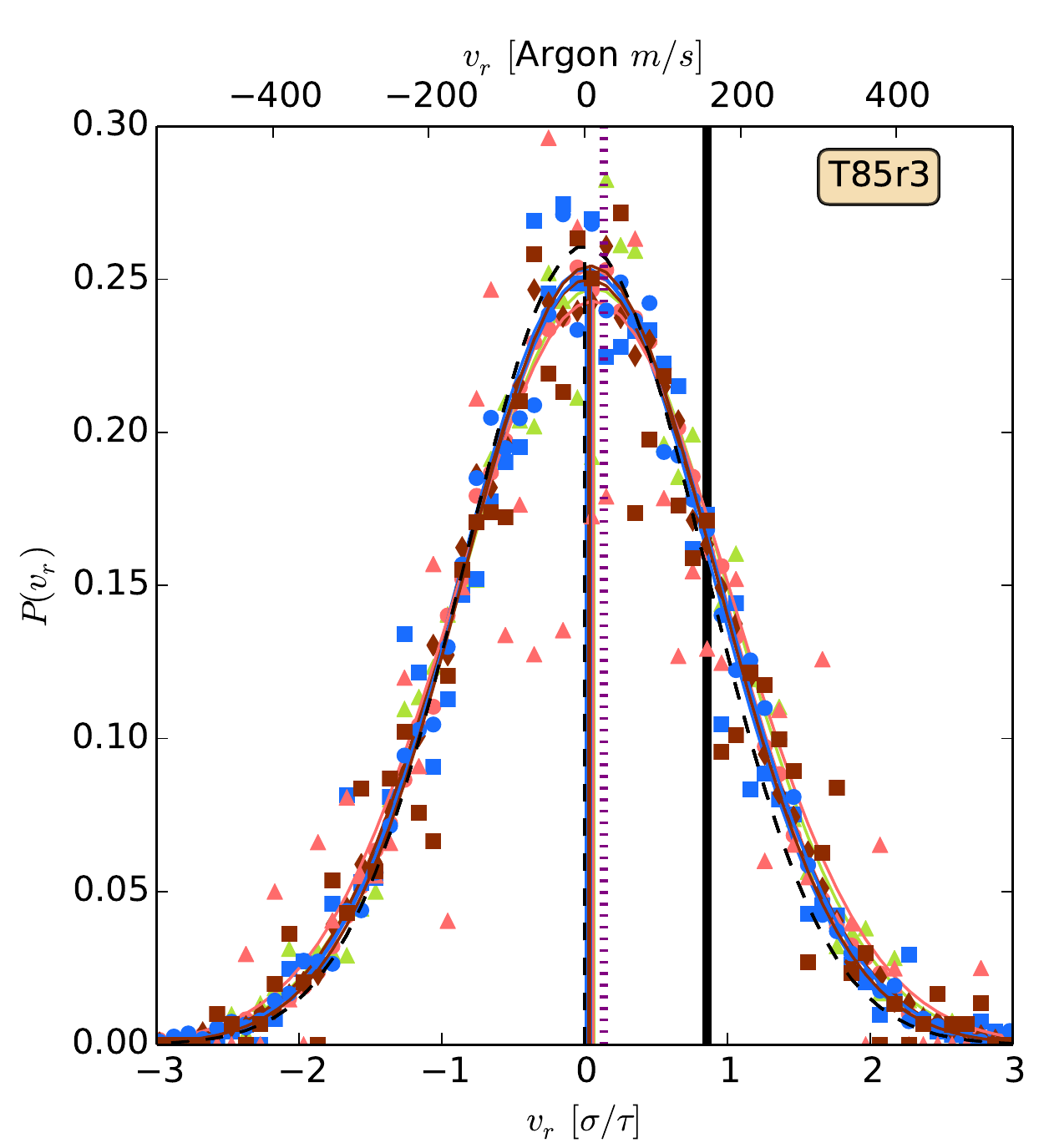}\caption{(Color online) Transition region radial velocity probability distributions for bubbles in run T85r3. Bubble growth is typically less rapid in the boiling than in the cavitation regime. The thick vertical black line indicates the sound speed in the liquid.}\label{fig:interface_rad_vel_t8}
\end{figure}

\begin{figure}
\includegraphics[scale = 0.5]{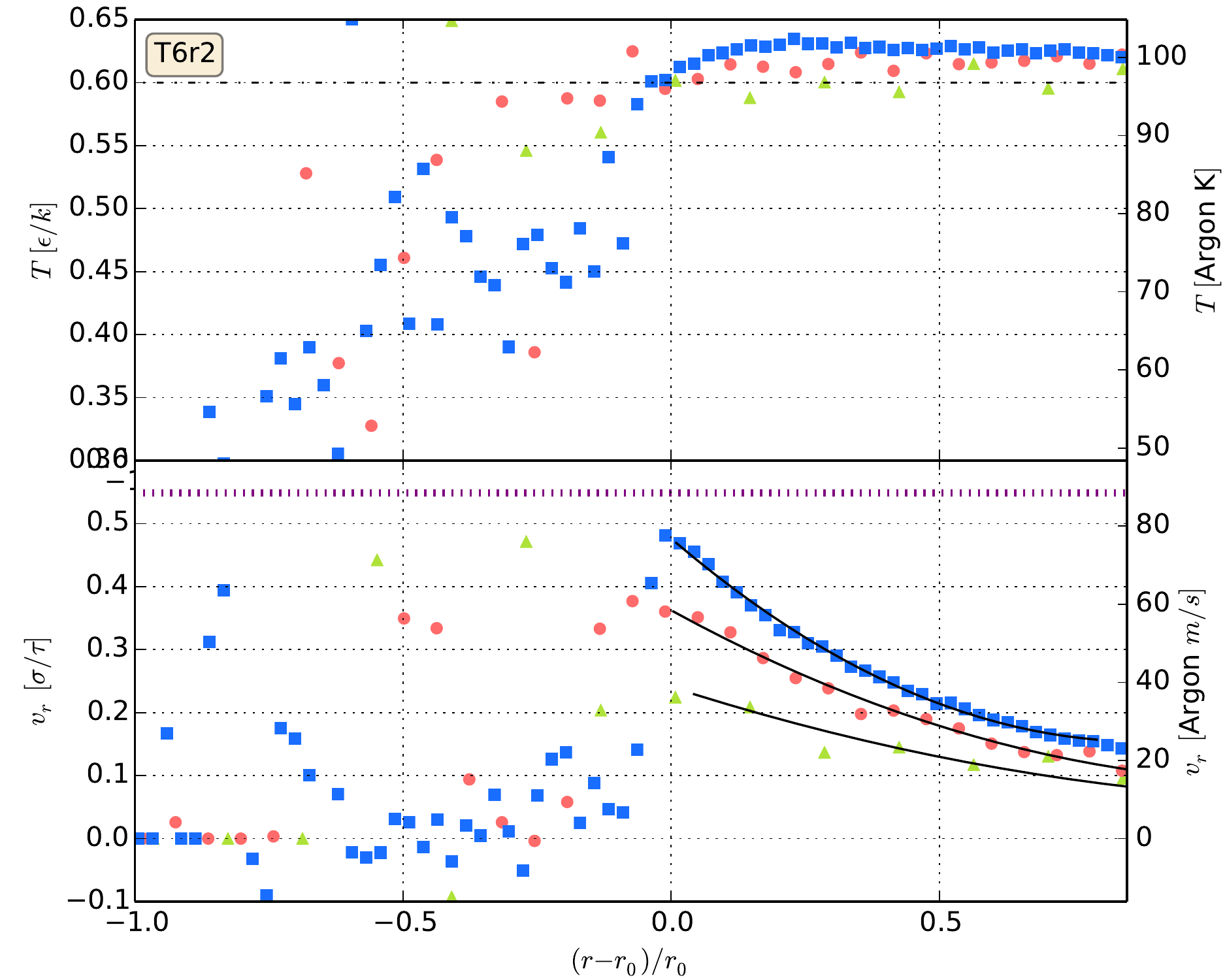}
\caption{(Color online) Temperatures (upper panel) and radial velocities (lower panel) for T6r2 bubbles. Per-bin temperatures have been calculated from per-bin radial velocity distributions. (An example of one such distribution is plotted in figure \ref{fig:interface_rad_vel_t6} for this run.) This bulk-motion adds an excess to the kinetic energies, which can be subtracted away to yield the temperature, plotted in the upper panel. The horizontal axis in both panels denotes the scaled-and-shifted radius, which centers all bubbles on top of one another. The purple horizontal line in the lower panel indicates the expected linear regime velocity. The radial velocity falloff deep into the liquid is parabolic, as expected from mass conservation\citep{brennen1995cavitation}. Best-fit parabolas are over-plotted with solid black lines.}\label{fig:temperature_profile_t6}
\end{figure}

\begin{figure}
\includegraphics[scale = 0.5]{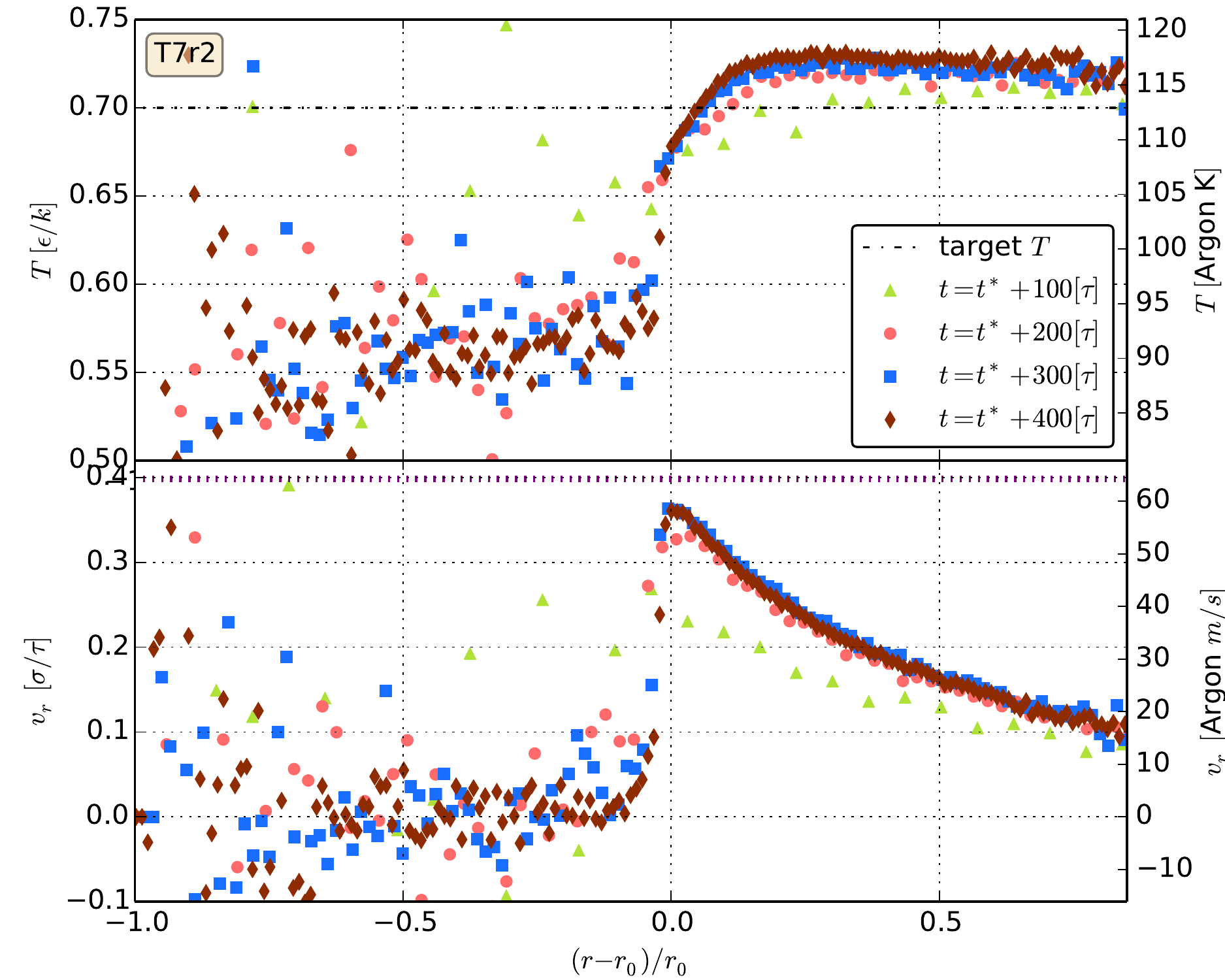}
\caption{(Color online) Temperatures (upper panel) and radial velocities (lower panel) for T7r2 bubbles.}
\label{fig:temperature_profile_t7}
\end{figure}

\begin{figure}
\includegraphics[scale = 0.5]{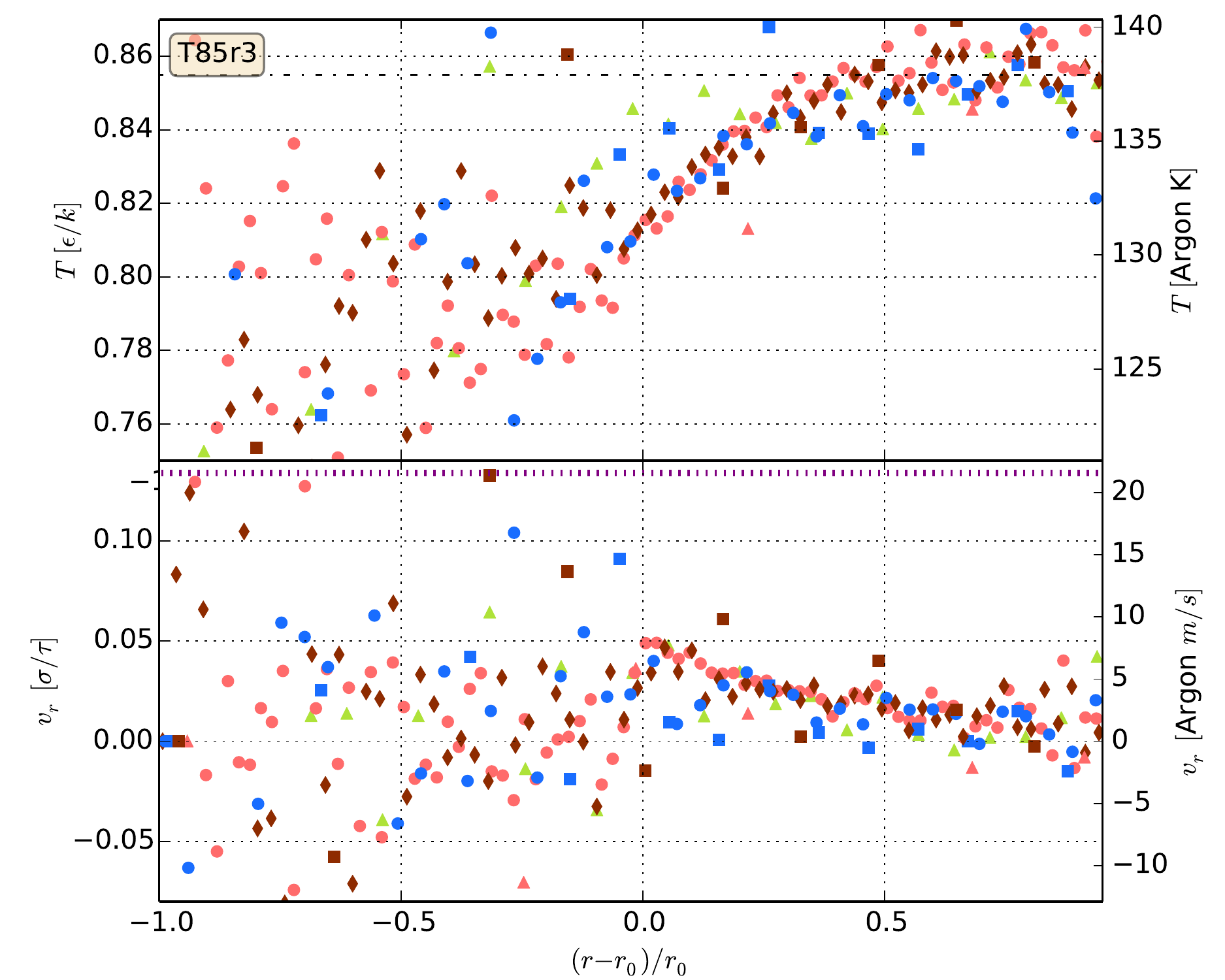}
\caption{(Color online) Temperatures (upper panel) and radial velocities (lower panel) for T85r3 bubbles.}\label{fig:temperature_profile_t85}
\end{figure}

\begin{figure}
\includegraphics[scale = 0.5]{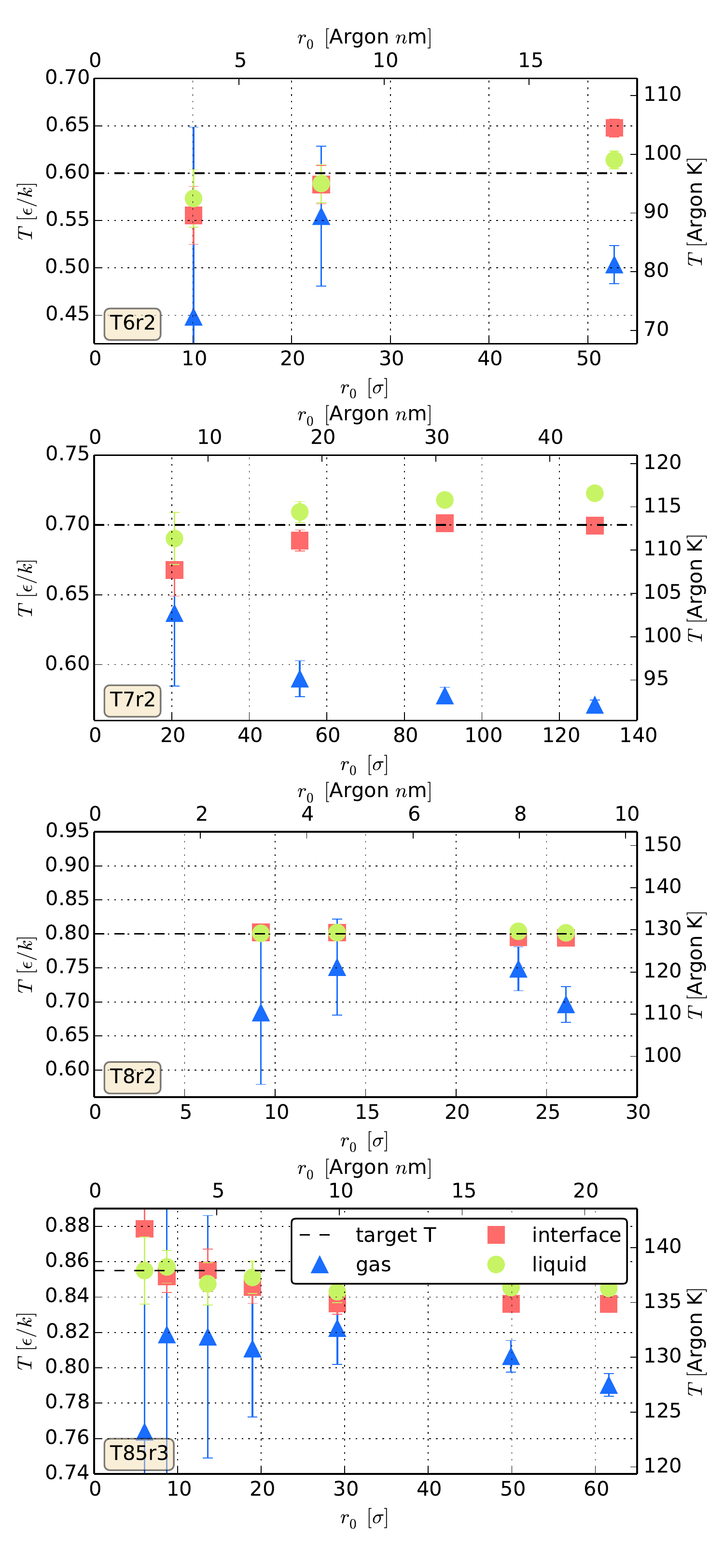}
\caption{(Color online) Temperatures within the gas, as well as over the interface. These are averages of radially binned temperature profiles (see figure \ref{fig:temperature_profile_t7} for example). We attribute the lower temperature in the gas to latent heat, and the increased temperature outside the bubble to the compression of the boundary in the fast bubble growth cases: T7r2 and T6r2. }\label{fig:T_vs_r0}
\end{figure}

The temperature of an ensemble of atoms is typically defined from their mean kinetic energy,
\begin{equation}
kT \equiv \frac{2}{3}\left<E_{\rm{kinetic, thermal}}\right> = \frac{1}{3N}\sum_{i=1}^{N}m_iv_i^2,
\end{equation}
where $N$ is the total number of atoms in the ensemble, $m_i$ are the atom masses, and the velocities $v_i$ are those relative to the simulation box. However, in the case of growing bubbles, which exhibit a preferred direction of motion we must take care to disentangle bulk particle motion from random thermal motion. For each bubble, for each radial bin, we convert Cartesian velocities to radial velocities. As examples, figures \ref{fig:interface_rad_vel_t6}-\ref{fig:interface_rad_vel_t8} plot the radial velocity probability distributions of the radial bin covering $r = r_0$, the midpoint of the transition for a few bubbles. From these we may determine the magnitude of the bulk radial motion in each radial bin. This artificial contribution to the temperatures may then be removed at each radial bin, yielding the temperature profile. For the same two runs, these are plotted against scaled bubble size in the upper panels of figures \ref{fig:temperature_profile_t6}-\ref{fig:temperature_profile_t85},  while the lower panels plot the radial velocity excess. Transition region and gas temperature averages are shown in figure \ref{fig:T_vs_r0}. 

In Diemand et al.\cite{bubble_paper_1}, we show that the initial formation (nucleation) of bubbles is an  isothermal process as the event is not preceded by above-average temperatures (local hot-spots). Here however, we demonstrate that bubble growth on the other hand is strongly non-isothermal, 
especially for cavitation runs with large negative liquid pressures. Specifically:

\begin{itemize}
\item Bulk bubble growth motions in the transition region and beyond is measurable from per-particle radial velocity averages. For the two runs at significant negative pressure, the bubble growth is rapid enough for this signal to be significantly higher than the noise, allowing for accurate bubble velocity measurements. These measurements agree with the velocities determined from the increase in the bubble volume. This velocity is close to that expected from the cavitation linear regime terminal speed (see section \ref{sec:bubble_growth}). The bulk motion contribution to the kinetic energies must be taken into account to calculate the temperature.  
\item For all bubbles, in all runs, we find that the gas temperatures are up to 20\% colder than the ambient run temperature. We observe a general trend of lower gas temperatures for increasing bubble sizes. This is expected from the latent heat used for evaporation.
\item For runs T85r3 and T8r2 the temperatures rise in the transition region and reach the run's average temperature around $1.5r_0$ from the bubble centers. 
\item T7r2's and T6r2's temperatures also rise in the transition region, but then they become significantly higher than the average simulation temperature (by $5\%$) just beyond $r_0$. They drop only slowly after this: one has to go out to about $4 r_0$ before the average simulation temperature is reached. This extra heat found around these bubbles is likely due to compression, which is significant in the cavitation regime when the bubble growth velocities are a significant fraction of the sound speed. 
\end{itemize}

\section{\label{sec:shapes}Shapes}

\begin{figure}
\includegraphics[scale = 0.5]{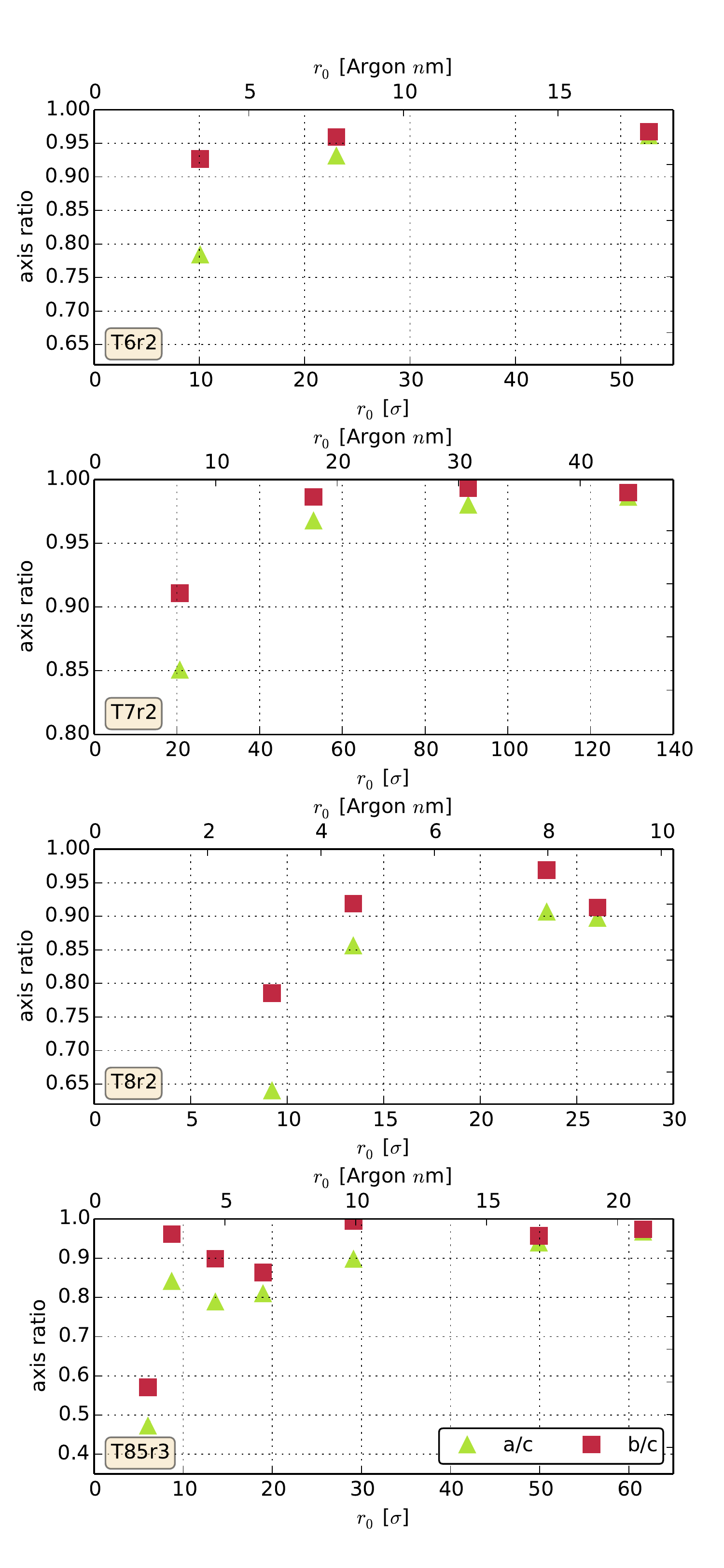}
\caption{(Color online) Bubble shapes: Axis ratios of unity correspond to sphericity, smaller values indicate more elongation. We observe a general trend of bubbles becoming more spherical the larger they are. For small sizes $r_0 <10\sigma$ the bubbles are far from spherical.}\label{fig:PCA}
\end{figure}

An assumption common to almost all nucleation and bubble evolution models is that the bubbles are spherical as this shape minimizes the surface energy. We measure the bubble shape by measuring their ellipsoid axes $a, b$ and $c$. This is done by principal component analysis (PCA) of all particles in the bubble and its vicinity with a coordination number of two or fewer. This results in the shape measurement being determined primarily by the inner part of the gas-liquid interface. Details on using PCA to determine shapes of particle distributions can be found in Zemp et al.\cite{2011ApJS..197...30Z} applied to dark matter distributions and Ang\'elil et al.\cite{2014JChPh.140g4303A} applied to atom clusters in gas-to-liquid MD simulations. Figure \ref{fig:PCA} plots the short-over-long $a/c$ and medium-over-long $b/c$ axis ratios against bubble size $r_0$.  

Non-spherical bubbles at small sizes was also reported in numerical simulations in \citep[Meadley and Escobedo, 2012]{escobedo}. 
We note that:
\begin{itemize}
\item The observed asphericity for smaller cluster sizes prompts caution for quantities measured using spherical bins. Artificial contributions to the quantity may enter due to atypical shapes at small sizes. See for example the measured interface thickness for the smallest bubbles in T7r2 and T85r2 (figure \ref{fig:number_density_profile} and the lower panel of figure \ref{fig:gas_and_d_vs_r0}).   
\item Bubbles become more spherical as they grow. At the critical size they are quite elongated with typical axis ratios of $a/c = 0.48$ and $b/c = 0.58$. Thermal fluctuations are likely responsible for this asphericity. This could have implications for nucleation, as such shapes have a higher-than-expected surface energy due to the increased surface area (preserving the volume), leading to a higher energy cost to form the critical bubble, and thereby lowering nucleation rates.  
\end{itemize}

\section{\label{sec:bubble_growth}Bubble Growth}

The Rayleigh-Plesset equation describes the evolution of the radius of a spherical bubble with a hydrodynamics description. The liquid is considered incompressible, and the interface sharp. The evolution equation is an equation for the acceleration of the interface radius, and is made up of various competing influences, some of which promote bubble growth, and others which squeeze the bubble. In this section we apply this description to our simulated bubbles, in order to test the hydrodynamics framework, as well as to determine the applicability of the model assumptions. 

The Rayleigh-Plesset equation reads\citep{OriginalRayleigh,RP2,plesset1949dynamics}
\begin{equation}\label{eq:RP}
R\ddot{R} = \underbrace{-\frac{3}{2}\dot{R}^2}_{\textrm{momentum}} + \frac{1}{m n_l}\left[\underbrace{P_g}_{\textrm{gas}}-\underbrace{P_l}_{\textrm{liquid}} - \underbrace{\frac{2\sigma}{R}}_{\textrm{surface}}-\underbrace{\frac{4\mu}{R}\dot{R}}_{\textrm{viscosity}} \right] 
\end{equation}
 where $R\left(t \right)$ (or short $R$) is the bubble radius at time $t$, $\dot{R}$ and $\ddot{R}$ its first and second time derivatives, $n_l$ is the liquid density (mostly constant over the time we consider), $P_i$ is the gas pressure in the bubble, $P_l$ is the pressure in the liquid, $\sigma$ the planar surface tension, and $\mu$ the viscosity.

Surface tension, viscosity and the pressure difference between the gas inside the bubble and the surrounding liquid compete with one another.

\begin{itemize}
\item \textbf{Surface tension}, $\sim R^{-2}$ works to squeeze the bubble closed. It plays an important role when the bubble is small, and less so as it grows due to its strong inverse radius dependence.
\item \textbf{Viscosity}, $\sim \dot{R}R^{-2}$, a frictional term, also acts to slow down the growth. Typically it reaches dominance after the surface tension term does, as bubble growth $\dot{R}$ picks up. As the bubble growth continues, the $\sim R^{-2}$ dependence sharply damps this effect.
\item \textbf{Internal pressure $P_g$} is always positive for a gas, and works to expand the bubble. This term is the one responsible for bubble growth in the boiling regime. In the cavitation regime it is usually irrelevant.   
\item \textbf{External pressure $P_l$} provides a $\sim R^{-1}$ effect. This is the only term which whose influence direction is not evident a priori. In the boiling regime ($P_l>0$) the ambient liquid pressure contributes to a sub-dominant squeezing, while in the cavitation ($P_l<0$) it is usually the most dominant expansion term. 
\end{itemize}

In our homogeneous liquid simulations, any bubble must start out small.
According to the classic nucleation theory, around the critical size (determined by the mechanical equilibrium condition $\dot{R} =0$), the surface tension and the liquid pressure difference are the most important contributions, and so other influences are neglected. These classical critical size estimates match our MD results very well\cite{bubble_paper_1}. In this paper we consider the two regimes following this: the intermediate growth regime - from the critical cluster size and onwards - into the linear growth regime. The linear regime is somewhat simpler as the effects of viscosity and surface tension become negligible, because of their steep inverse radius dependence. 

\subsection{Linear regime growth\label{sec:lin_growth}}

Many molecular dynamics approaches to testing the Rayleigh-Plesset equation as a description for bubble evolution disregard the effects of surface tension and viscosity, valid when the bubbles are large. See MD simulations for large bubble collapse \citep{RP_verification}. The largest bubbles in our simulations enter the same regime towards the ends of the runs. We can compare the large-bubble Rayleigh-Plesset regime to the bubble growth velocities which we measure in simulations.

\begin{itemize}
\item \textbf{Liquid pressure-dominated} In this regime the bubble growth is rapid due to strong negative pressure, and the Rayleigh-Plesset equation \eqref{eq:RP} reduces to 
\begin{equation}\label{eq:final_explosion}
\dot{R} = \sqrt{-\frac{2}{3}\frac{P_l}{m n_l}}.
\end{equation}
This is typical for cavitation, where the bubble gas pressure is a less dominant source of growth than the strong negative liquid pressure.

\item \textbf{Boiling regime} means the Rayleigh-Plesset equation becomes
\begin{equation}\label{eq:final_ME}
\dot{R} = \sqrt{\frac{2}{3}\frac{P_g - P_l}{m n_l}},
\end{equation}
where $P_g$ is the current pressure in the bubble. 
\end{itemize}

\begin{figure}
\includegraphics[scale=0.55]{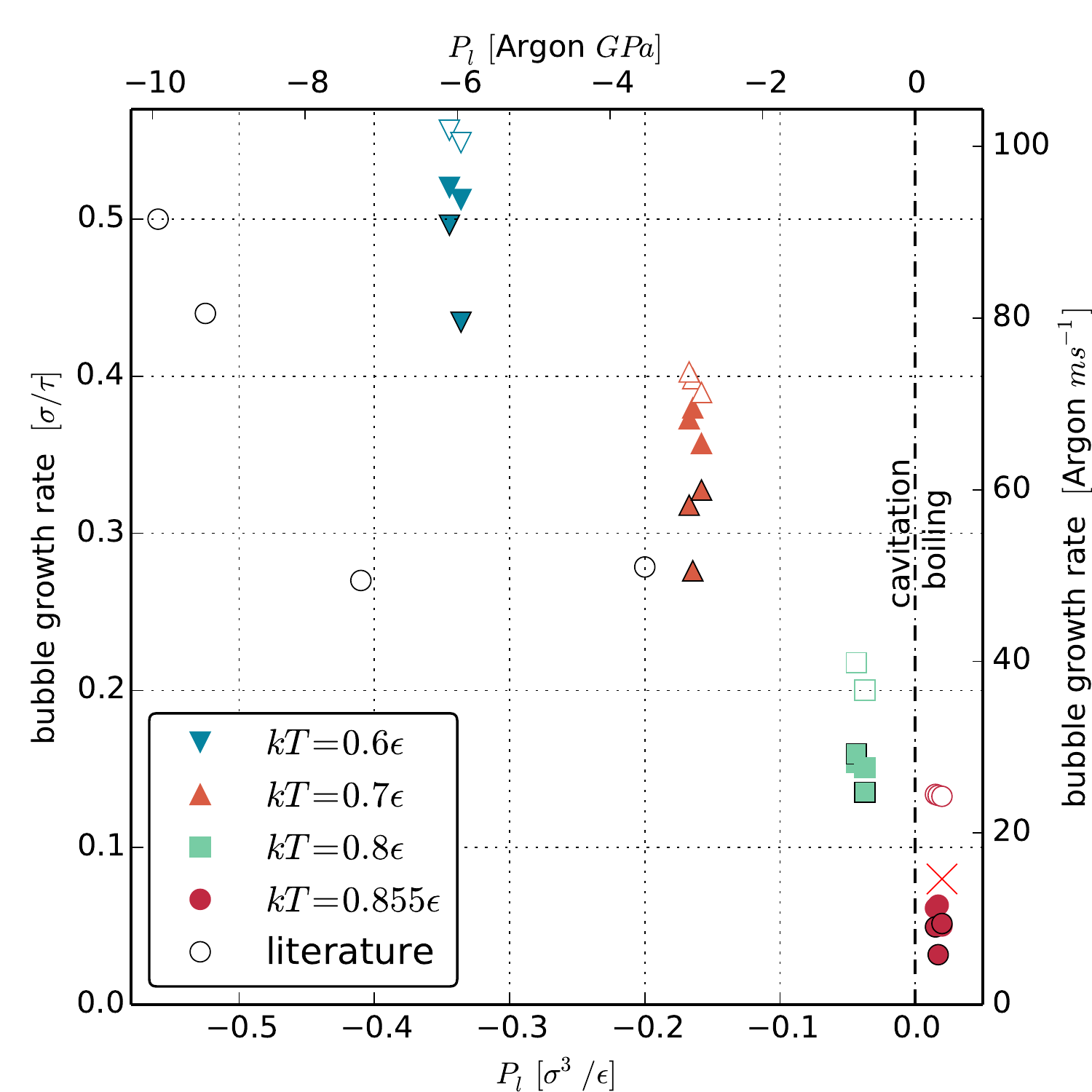}
\caption{(Color online) Late-stage bubble growth velocities for the largest two bubbles in each run. The largest is indicated by the solid markers, the second largest by black-rimmed edge markers. The hollow markers denote the expected terminal velocity values, equations \eqref{eq:final_explosion} and \eqref{eq:final_ME}. The single `$\times$' marker indicates the expected linear growth velocity due to a non-isothermal equation of state, and an evaporation efficiency $\alpha = 0.4$. We suspect that our boiling regime bubbles have not yet reached the linear growth stage, and are still accelerating. The first three literature markers are from Kuksin et al. (2010). However their smaller bubbles and somewhat higher surface tensions (due to a Lennard-Jones potential without a force shift) may not have reached the linear regime and so are still accelerating at the time of measurement. The fourth literature value is from Wu (2003)\citep{wu_pan}, however their small computational volumes meant that bubble genesis and growth effects an immediate liquid pressure increase, which quickly suppresses the growth rate. This measurement therefore may be regarded as a lower bound. Similarly, Kinjo (1998)\cite{Kinjo1998343} reports bubble growth rates $\sim 0.3\sigma/\tau$, yet at pressures $<-0.7 \sigma^3/\epsilon$ - far beyond the regime we probe in our simulations.}\label{fig:linear_velocities}
\end{figure}

The dotted purple lines in radial velocity plots, in figures \ref{fig:temperature_profile_t6}-\ref{fig:temperature_profile_t85} show these expected maximum velocities. Figure \ref{fig:linear_velocities} uses histogram bubble data (explained in section \ref{sec:numerical_simulations}) at the end of the run to evaluate the rate at which the largest two bubbles in each run grow. 

We find that
\begin{itemize}
\item For the boiling regime runs at $kT=0.855\epsilon,$ the measured late stage bubble growth is half the prediction from \eqref{eq:final_ME}. We attribute this to uncertainties in the bubble gas pressure $P_g$. We have assumed the ideal gas law, and assumed that the temperature is at the reduced temperature (see first panel in figure \ref{fig:T_vs_r0}), with the gas number density that reported in figure \ref{fig:number_density_profile}. We suspect that even the largest bubbles at this temperature were not yet in the linear regime. This is confirmed by the full Rayleigh-Plesset solution for run T85r3, which we perform in the following subsection.  
\item If we replace the gas pressure in \eqref{eq:final_ME} with a non-isothermal equation of state - using the expected gas number density in the linear regime to find the pressure assuming a temperature not of $kT=0.855\epsilon$ but rather $kT=0.8\epsilon$ (as shown in the lowest panel of Figure \ref{fig:T_vs_r0}), and choose an evaporation efficiency (see equation \eqref{eq:bk}) of $\alpha = 0.4$, then we find an expected linear growth rate somewhat closer to the simulation measurement. When we decrease the bubble gas pressure due to the lowered temperature, the diffusion occurs more rapidly, and so the efficiency also needs to be lowered so as to obtain a result which matches with the simulation measurements for $\dot{R}$. On Figure \ref{fig:linear_velocities} we have marked this expected value for the run T85r3 with a `$\times$' marker. 
\item Predictions for negative-pressure dominated bubble growth are more accurate. Because the pressure of the liquid increases slightly over the entire simulation, which would slow down the bubbles down, linear regime growth rate predictions from \eqref{eq:final_explosion} should be taken as upper bounds, as we input the initial liquid pressure (As given in table \ref{tab:details}). We find that the simulation bubbles typically grow $\sim 10\%$ slower than the Rayleigh-Plesset negative-liquid-pressure dominated expectation. 
\end{itemize}

\subsection{Intermediate growth regime: the RP-diffusion system}

In this section we test the capability of the Rayleigh-Plesset equation to predict bubble size evolution from the critical size, on to the end of the simulation. We may not make any size-approximations in this regime. We assume that the Rayleigh-Plesset thermodynamic parameters are taken at the simulation's average temperature, disregarding the (as we shall show, likely important) complications which arise from the non-isothermal profiles. (See figure \ref{fig:interface_rad_vel_t8} for example.)  

The Rayleigh-Plesset equation depends on the pressure inside the bubble, $P_g$, which is time-dependent, not known a priori, and difficult to measure directly in simulations. To get the gas pressure, we must model the diffusion process through the bubble wall. The net rate of evaporation into the bubble is driven by the difference between gas pressure $P_g$ and its vapor pressure $P_v\left(kT \right)$. 
The time derivative of the number of molecules in the bubble $N\left(t\right)$ is provided by the Hertz-Knudsen relation\citep{knudsen}

\begin{equation}\label{eq:bk}
\dot{N}\left(t\right) = \frac{4\pi \alpha R\left(t\right)^2}{\sqrt{2\pi m k T}}\left[P_v - P_g \right],
\end{equation}
where $\alpha$ is an evaporation efficiency. The efficiency is analogous to the sticking probability or condensation efficiency measured from MD growth curves of liquid-like clusters\cite{Tanaka2005,Tanaka2011,paper1}. However unlike the condensation efficiency, the evaporation coefficient $\alpha$ may be greater than unity. This is supported by theoretical arguments and experiments\citep{holyst3}, as well as recent Lennard-Jones numerical simulations of liquid droplet\citep{holyst1} and liquid slab evaporation\citep{holyst2}. Along with an equation of state, the Hertz-Knudsen equation provides the gas pressure, which appears also in the Rayleigh-Plesset equation \eqref{eq:RP}. Together, equations \eqref{eq:RP} and \eqref{eq:bk} form a coupled system which we will solve simultaneously. 

The vapor pressure in the bubble differs from the equilibrium vapor pressure because the liquid pressure is less than its equilibrium pressure. The relevant shift is called the Poynting correction, and is
 \begin{equation}
P_v = P_e\delta + P_l\left(1-\delta\right),
 \end{equation}
which depends on the liquid-gas density ratio,
\begin{equation}
\delta = 1 - \frac{n_g}{n_l} + \frac{1}{2}\left(\frac{n_g}{n_l}\right)^2.
\end{equation}
This amounts to a significant correction of $10\%$ for our $kT=0.855$ simulations. See table \ref{tab:details} for the values of the thermodynamics parameters which we use. Although techniques exist for directly measuring the pressure within the bubble $P_g$, (using the Kirkwood-Irving pressure tensor\citep{thompson}) we are unable to do so due to low signal-to-noise levels. Instead we work with the number density within the bubble $n_g$. To convert the bubble gas density $n_g$ into $P_g$ - which appears both in \eqref{eq:RP} and \eqref{eq:bk}, we require an equation of state. Either we
\begin{enumerate}
\item assume an ideal gas, (as we shall do for T7r2) replacing $P_g\left(t\right)$ with
\begin{equation}
P_i\left(t\right) = \frac{3 N\left(t\right)kT }{4\pi R\left(t \right)^3}, 
\end{equation}
or
\item use a simulation-based equation of state estimate. We perform $2.5\sigma$ TSF-LJ gas simulations at the bubble gas temperatures and number densities, and measure the bulk gas pressure. We may further implement non-isothermality by lowering the $P_g$ by the temperature ratio. As we shall show, this is a necessary addition to the Rayleigh-Plesset model to accurately reproduce bubble growth for T85r3. 
\end{enumerate}

\begin{figure}
\includegraphics[scale = 0.6]{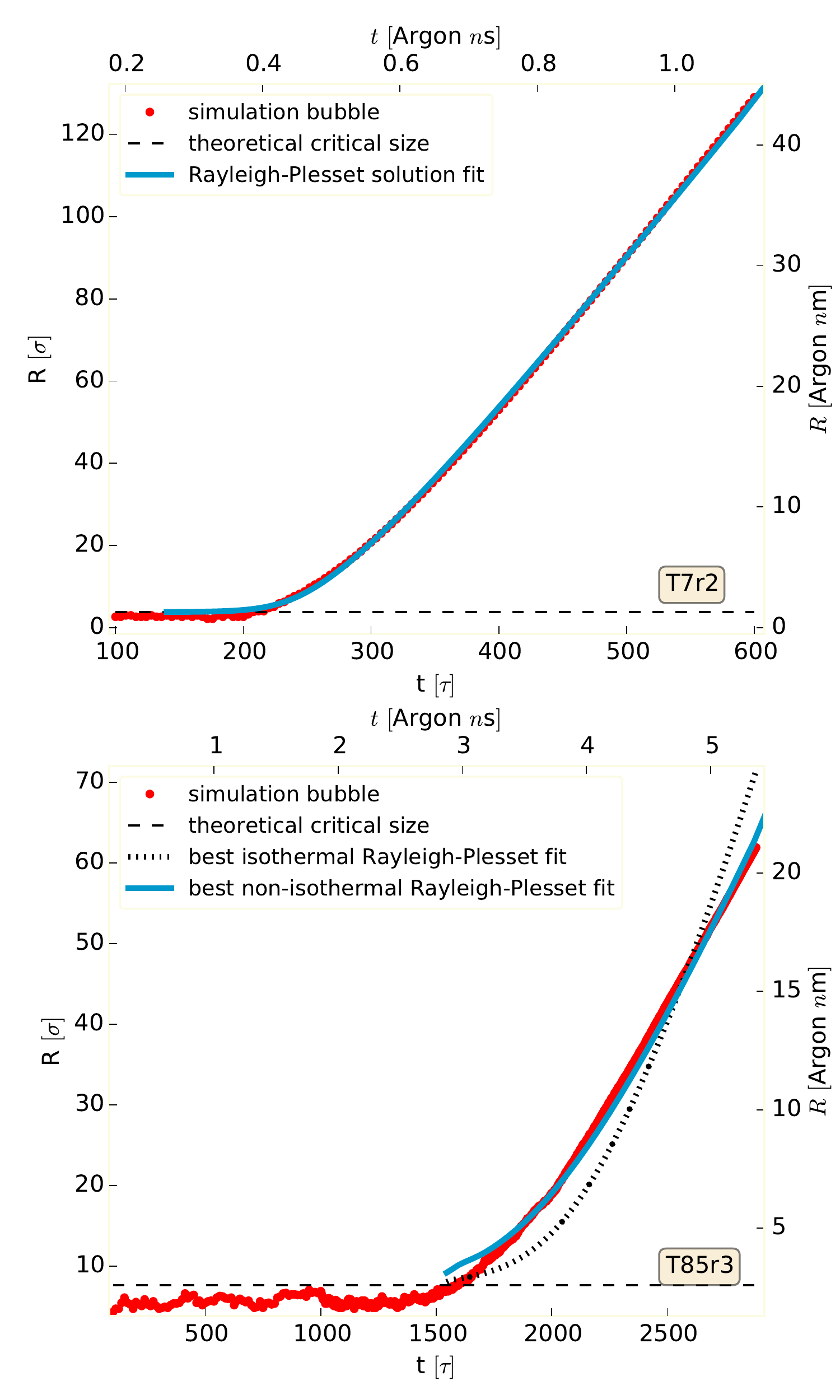}\caption{(Color online) Simulation bubble sizes measured (red markers) for the largest bubbles in runs T7r2 and T85r3. The solid lines are the Rayleigh-Plesset model fits. The accompanying bubble growth rates and the bubble gas densities (all three are coupled) solutions are in the first two rows of \ref{fig:RP_detail}. With the isothermal model, the cavitation regime fit, T7r2 matches well, and the boiling regime somewhat poorly, predicting in particular a different linear growth rate than the one we measure. However, we get a good fit when using a non-isothermal gas pressure estimate.}
\label{fig:RP_r}
\end{figure}

\begin{figure*}
\includegraphics[scale = 0.5]{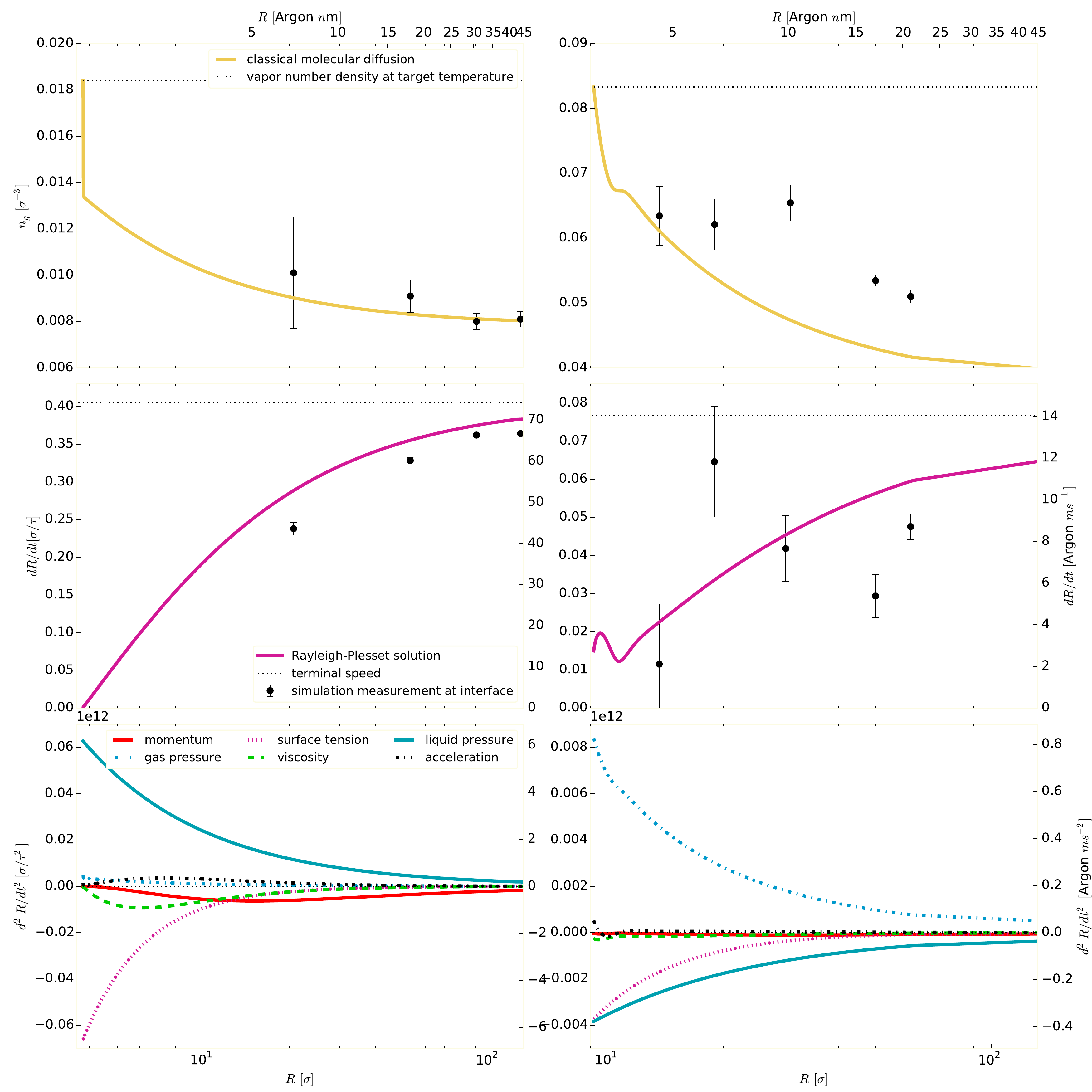}\caption{(Color online) Top row: gas densities solutions, equation \eqref{eq:bk}, for the coupled RP-diffusion system. The fit via viscosity $\mu$ and $\alpha$ is done for the bubble sizes (red dots in figure \ref{fig:RP_r}). The black markers are the final densities measured in section \ref{sec:interfaces} (black markers). Middle row: growth rate solutions. The markers are the measured radial velocity profile excesses, detailed in section \ref{sec:temperatures}; and show good agreement with the Rayleigh-Plesset solution. Notice that the RP solution does not quite reach the linear growth regime, especially for T85r3. This means that the linear-regime growth estimates in section \ref{sec:lin_growth} should be taken as upper bounds when comparing to the bubble growth rates at the end of each run. Bottom: the acceleration breakup of the competing influences which make up the RP model. In other words, the right-hand-side of equation \eqref{eq:RP}. These panels illustrate the contrast in bubble growth influence factors in the boiling vs. cavitation regimes. The black line is the total acceleration - the sum of the others.} 
\label{fig:RP_detail}
\end{figure*}

The Rayleigh-Plesset equation and the diffusion equation form a coupled system of ordinary differential equations.  We solve the RP-diffusion equation system as a shooting problem, using an optimizer to adjust the viscosity. We fit the Rayleigh-Plesset solution to the bubble sizes (red markers in figure \ref{fig:RP_r}), as well as to the final bubble gas density measurements. We do this for the largest bubbles in runs T7r2 and T85r3. The results for the predicted sizes are shown as solid blue curves in figure \ref{fig:RP_r}, and the bubble growth rate, and gas densities in figure \ref{fig:RP_detail}.

\begin{enumerate}
\item \textbf{T7r2} Here, we use an ideal gas equation of state, and treat the system as isothermal.  The best fit viscosity for this bubble growth solution is $\mu = 0.58 \pm 0.01$ and efficiency $\alpha = 1.69\pm 0.34$. We note that in the cavitation regime, this value for the efficiency is not constrained by the bubble growth $R\left(t\right)$ measurements (upper panel figure \ref{fig:RP_r}), but by bubble gas density measurements (upper left panel figure \ref{fig:RP_detail}). This is because the Hertz-Knudsen equation \eqref{eq:bk} couples weakly to the RP equation due to the bubble gas pressure playing a sub-dominant role in bubble growth (see the blue curve on the lower left panel of figure \ref{fig:RP_detail}). 

\item \textbf{T85r3} For the largest bubble in T85r3, no isothermal model is able to match the bubble evolution (see the black line in figure \ref{fig:RP_r}). Implementing a non-isothermal equation of state however, which lowers the pressure proportional to the measured bubble gas temperature (see section \ref{sec:temperatures}) gives a good fit, with best fit solution $\mu = 0.52\pm 0.08$ and $\alpha = 0.41\pm 0.03$. 
\end{enumerate}

In both cases, the best fit viscosity values seem reasonable in comparison with the range of values
predicted by MD viscosity measurement simulations\citep{woodstock,Rowley1997,Muller1999,Meier2004,Galliero2005} for LJ liquids with cutoff scales of $2.5\sigma$ and larger.
We did not find viscosity values for our TSF-LJ liquid, however the viscosities are quite robust to variations around the LJ potential from Galli{\'e}ro et al\cite{Galliero2005}.
Classical models implicitly assume an evaporation efficiency of unity\cite{blander_katz}. However, Lennard-Jones numerical simulations of droplet evaporation support the possibility of larger efficiencies\citep{holyst1}, as we have found in the cavitation regime case fits, with $\alpha = 1.69\pm 0.34$. In the boiling regime we find a significantly smaller efficiency $\alpha = 0.41\pm 0.03$. Low efficiencies have also been derived from growth curves of liquid-like clusters in homogeneous vapor-to-liquid nucleation simulations\cite{Tanaka2011,paper1}.

The fits are successful, however we caution that the system is overdetermined because of the simplicity of the growth curves. We have attempted to add sophistication to the RP model by including a Tolman length\citep{tolman} $\delta$ into the surface tension, (promoting $\sigma\rightarrow\sigma = \sigma_0-2\delta/R$) as part of the parameters-to-be fitted, since a small, positive $\delta$ seems to better fit droplet evaporation\cite{holyst4} rates and bubble nucleation rates\cite{paper3,Kuksin2010,Baidakov2014,bubble_paper_1}. However the system quickly becomes overdetermined as the correlation between the fit parameters is close to unity, leading to inconclusive results.

Summarizing, we find that:
\begin{itemize}
\item While in the example solved here, we have set the initial value for the gas density $n_g$ to the gas equilibrium density, the final value for the gas density $n_g$ is largely unaffected by its initial value. During the evolution, density drops below the vapor pressure at the simulation average temperature, likely because of the reduced vapor pressure due to the reduced local temperature. 
 
\item Bubble growth in the cavitation and boiling regimes are remarkably different from one another, as shown in the 3rd row of figure \ref{fig:RP_detail}. In both cases almost all terms in the Rayleigh-Plesset model play an important role. The difference between the two regimes is when they come into play, and the different phases of dominance that the individual terms have. For example, when the bubbles are small, in the cavitation regime, the acceleration due to the liquid pressure is balanced by the surface tension and the viscosity which decelerate the growth. At size $\sim 12\sigma,$ the momentum, viscosity and surface tension terms are all contribute equally to balance the growth due to the liquid pressure. In the boiling regime case, the acceleration is caused by the bubble gas pressure term, and is balanced by the surface tension and liquid pressure which work to slow down the growth. 

\item For the boiling regime case, where bubble acceleration is dominated by diffusion into the bubble due to the vapor pressure - gas pressure difference \eqref{eq:bk}, the addition of a non-isothermal gas pressure is necessary because the latent heat produced by the transformation has lowered the temperature. This lower gas pressure increases the rate at which diffusion into the bubble occurs. This is then balanced by an evaporation efficiency $\alpha = 0.4,$ in order to produce the correct behavior. 
\end{itemize}

Our implementation of the Rayleigh-Plesset bubble growth model has weaknesses:
\begin{itemize}
\item While we have attempted to include a non-isothermal bubble gas pressure, there are other ways non-isothermal effects would enter in a more complete description. Thermodynamic quantities such as viscosity, and planar surface tension are not at the run's average temperature. A more complete heat diffusion treatment which includes latent heat, compressive heat, and convective effects and their couplings to the fluid mechanics of bubble growth, as well as their effects on the Hertz-Knudsen evaporation relation would be a more realistic\citep{brennen1995cavitation, plesset_heat_transfer, WilhelsmenHeat, holyst3, holyst5}, albeit challenging approach. 
\item The compression of the liquid outside the bubble in T7r2 (discussed in section \ref{sec:interfaces}) is likely significant enough to alter the viscosity of the fluid directly outside the bubble, not only because of the raised temperature, but also the raised density\citep{woodstock,Rowley1997,Muller1999,Meier2004,Galliero2005}. The inclusion of a more intricate viscosity profile may marginally improve the model accuracy when the bubble growth is rapid, typical for the cavitation regime.  
\item The assumption of constant liquid pressure may hurt the late-time RP evolution. In most runs the pressure is raised by a few percent by the end of the simulation. Replacing $P_l$ with a time-dependent liquid pressure may improve the evolution prediction late in the run. 
\item The Rayleigh-Plesset model assumes a Heaviside-like sharp density profile over the bubble interface. Simulation and experimental bubbles have interfaces often of order their sizes. A more realistic model would attempt to provide not $R=R\left(t\right)$ but $n=\left(R,t\right)$ - i.e. the evolution not of the radius, but the density profile.
\item At small sizes where the surface tension is important, and the curvature of the interface significant, a Tolman-like\citep{tolman} correction to the surface tension may be more realistic\cite{paper3,Kuksin2010,Baidakov2014,bubble_paper_1, holyst4}, and could be included in the surface tension in the Rayleigh-Plesset equation. Similarly, curvature could affect the vapor pressure of small droplets\citep{holyst3,holyst4}.
\item The RP model implementation relies on knowing the bubble gas pressure, entering both in the Rayleigh-Plesset as well as in the Hertz-Knudsen equation. For small bubbles this fluid is an intermediate phase, whose equation of state is unknown. A better understanding of this fluid, such as numerical simulations which measure pressure tensors directly for these small bubbles, would improve the model accuracy. Alternatively, in the boiling regime, by using energy conservation it may be possible to relate the bubble growth to energy flux, and use the Hertz-Knudsen relation to determine the bubble gas pressure\citep{holyst1}.  
\end{itemize}

\section{Conclusions}

Our simulations have allowed us to measure the properties of bubbles in $2.5\sigma$ cutoff TSF-LJ simulations. An understanding of the bubble properties will help to identify weaknesses in bubble nucleation and evolution models. We summarize our most interesting findings below.
\begin{itemize}
\item Fast growing bubbles (typically in the cavitation regime) create a shock-wave in the liquid, evident due to a $5\%$ over-density in the liquid surrounding the bubble. This induces compression heating of the surrounding liquid, also up to $5\%$.
\item Bubble gas temperatures are $20\%$ lower than the ambient run temperatures. They continue to drop as the bubbles grow. This internal temperature drop is due to the latent heat of transformation.  
\item Bubble gas densities are lower than expected from bulk simulations at the simulation average temperature, yet do agree with the densities expected from bulk simulations at the reduced temperatures.
\item Liquid-gas interfacial transition regions are measured to be $50-80\%$ thinner than measured from slab coexistence simulations at the target temperature.
They agree well with slab predictions for the lower temperatures found in the gas and the interface. 
\item We measure a critically sized bubble to have axis ratios $a/c = 0.48$ and $b/c = 0.58$, which correspond to an ellipsoidal shape, contrary to the spherical assumption used by most nucleation models. The bubbles quickly become spherical as they grow.
\item The Rayleigh-Plesset bubble growth model exhibits good agreement with the linear regime growth rates (within $20\%$) in the cavitation regime. For the boiling regime, the comparison is more difficult. The non-isothermality affects the diffusion rate into the bubble (the primary growth force), making computing the linear growth velocities more involved. Furthermore, our late-stage boiling regime bubbles are not yet in the linear growth regime.
\item The Rayleigh-Plesset bubble growth model is able to correctly match the intermediate behavior, where surface tension and viscosity are important for bubble growth, when we fit for viscosity. The best-fit shear viscosity values are consistent with those measured in dedicated MD simulations\citep{woodstock, viscos1}. Using a non-isothermal equation of state, in the boiling regime, the best-fit evaporation efficiency is $\alpha = 0.41\pm0.03$. For the cavitation regime case which we analyse, we find a best-fit evaporation efficiency greater than unity, $\alpha=1.69\pm 0.34$.
\end{itemize}

\begin{acknowledgments}
Computations were performed on SuperMUC at LRZ, on Rosa at CSCS, and on the zBox4 at UZH. J.D. and R.A. are supported by the Swiss National Science Foundation. 
\end{acknowledgments}

\bibliographystyle{apsrev4-1}
\bibliography{paper_2}

\end{document}